\documentclass[a4paper,fleqn,usenatbib,letters]{mnras}

\usepackage{newtxtext,newtxmath}

\usepackage[T1]{fontenc}
\usepackage{ae,aecompl}

\pdfoutput=1
\usepackage{natbib}
\usepackage{graphicx}
\usepackage{url}
\usepackage{microtype}
\usepackage{float}
\usepackage{afterpage}
\usepackage{longtable}
\usepackage[bottom]{footmisc}
\usepackage{xcolor}

\newcommand{\hi}{H{\sc i}}

\newcommand{\co}{CO(1--0)}

\makeatletter
\renewcommand*{\@fnsymbol}[1]{\star}
\makeatother

\title[VALES: III. Dust--gas calibration]{VALES: III. The calibration between the dust continuum and interstellar gas content of star-forming galaxies}

\author[T. M. Hughes et al.]{T. M. Hughes$^{1\thanks{E-mail: thomas.hughes@uv.cl}}$, 
E. Ibar$^{1}$, 
V.~Villanueva$^{1}$,
M.~Aravena$^{2}$, 
M.~Baes$^{3}$, 
N.~Bourne$^{4}$,\newauthor
A.~Cooray$^{5}$, 
L.~J.~M.~Davies$^{6}$,
S.~Driver$^{6,7}$,
L.~Dunne$^{4,8}$, 
S.~Dye$^{9}$,
S.~Eales$^{8}$,\newauthor
C.~Furlanetto$^{9,10}$, 
R.~Herrera-Camus$^{11}$, 
R.~J.~Ivison$^{4,12}$, 
E.~van~Kampen$^{12}$, \newauthor
M.~A.~Lara-L\'{o}pez$^{13}$,
S.~Maddox$^{4,8}$,
M.~J.~Micha{\l}owski$^{4}$, 
I.~Oteo$^{4,12}$, 
D.~Smith$^{14}$,\newauthor
M.~W.~L.~Smith$^{8}$,
E.~Valiante$^{8}$,  
P.~van~der~Werf$^{15}$, 
S. Viaene$^{3,14}$,
Y.~Q.~Xue$^{16}$
\\
$^{1}$Instituto de F\'{i}sica y Astronom\'{i}a, Universidad de Valpara\'{i}so, Avda. Gran Breta\~{n}a 1111, Valpara\'{i}so, Chile\\
$^{2}$N\'{u}cleo de Astronom\'{i}a, Facultad de Ingenier\'{i}a, Universidad Diego Portales, Av. Ej\'{e}rcito 441, Santiago, Chile\\
$^{3}$Sterrenkundig Observatorium, Universiteit Gent, Krijgslaan 281-S9, Gent 9000, Belgium\\
$^{4}$Institute for Astronomy, University of Edinburgh, Royal Observatory, Edinburgh EH9 3HJ, UK\\
$^{5}$Department of Physics and Astronomy, University of California, Irvine, CA 92697, USA\\
$^{6}$International Centre for Radio Astronomy Research, University of Western Australia, Crawley WA 6009, Australia\\
$^{7}$School of Physics and Astronomy, University of St Andrews, North Haugh, St Andrews KY16 9SS, UK\\
$^{8}$School of Physics and Astronomy, Cardiff University, The Parade, Cardiff CF24 3AA, UK\\
$^{9}$School of Physics and Astronomy, University of Nottingham, University Park, Nottingham NG7 2RD, UK\\
$^{10}$CAPES Foundation, Ministry of Education of Brazil, Bras\'{i}lia/DF 70040-020, Brazil\\
$^{11}$Max-Planck-Institut f{\"u}r extraterrestrische Physik, Giessenbachstra{\ss}e, 85748 Garching, Germany\\
$^{12}$European Southern Observatory, Karl-Schwarzschild-Strasse 2, 85748, Garching, Germany\\
$^{13}$Instituto de Astronom\'{i}a, Universidad Nacional Autonoma de M\'{e}xico, A.P. 70-264, 04510 M\'{e}xico, D.F., M\'{e}xico\\
$^{14}$ Centre for Astrophysics Research, University of Hertfordshire, Hatfield, Hertfordshire AL10 9AB, UK \\
$^{15}$Leiden Observatory, Leiden University, PO Box 9513, 2300 RA Leiden, The Netherlands\\
$^{16}$CAS Key Laboratory for Researches in Galaxies and Cosmology, Center for Astrophysics, Department of Astronomy, \\
$^{\ \ }$University of Science and Technology of China, Chinese Academy of Sciences, Hefei, Anhui 230026, China \\
}

\date{Accepted 2017 February 23. Received 2017 February 22; in original form 2017 February 18.}

\pubyear{2017}

\begin{document}
\label{firstpage}
\pagerange{\pageref{firstpage}--\pageref{lastpage}}

\maketitle

\begin{abstract}
We present the calibration between the dust continuum luminosity and interstellar gas content obtained from the Valpara\'{i}so ALMA Line Emission Survey (VALES) sample of 67 main-sequence star-forming galaxies at \mbox{$0.02<z<0.35$.} We use \co \ observations from the Atacama Large Millimetre/submillimetre Array (ALMA) to trace the molecular gas mass, $M_{\mathrm{H}_{2}}$, and estimate the rest-frame monochromatic luminosity at 850~$\mu$m, $L_{\nu_{850}}$, by extrapolating the dust continuum from MAGPHYS modelling of the far-ultraviolet to submillimetre spectral energy distribution sampled by the Galaxy And Mass Assembly (GAMA) survey. Adopting $\alpha_{\rm CO}$\,=\,6.5\,(K\,km\,s$^{-1}$\,pc$^{2}$)$^{-1}$, the average ratio of \mbox{$L_{\nu_{850}}$/$M_{\mathrm{H}_{2}}$~$=$~(6.4$\pm$1.4)$\times$10$^{19}$~erg~s$^{-1}$~Hz$^{-1}$~M$_{\odot}^{-1}$,} in excellent agreement with literature values. We obtain a linear fit of\mbox{ $\log_{10} \left({M}_{\mathrm{H}_{2}}/{\mathrm{M}_{\odot}}\right) =  (0.92\pm0.02) \log_{10} (L_{\nu_{850}}/{\mathrm{erg}~\mathrm{s}^{-1}~\mathrm{Hz}^{-1}})-(17.31\pm0.59)$.} We provide relations between $L_{\nu_{850}}$, $M_{\mathrm{H}_{2}}$ and $M_{\mathrm{ISM}}$ when combining the VALES and literature samples, and \mbox{adopting a Galactic $\alpha_{\rm CO}$ value}.
\end{abstract}
\begin{keywords}
galaxies: ISM -- ISM: lines and bands -- submillimetre: galaxies
\end{keywords}

\section{Introduction}

Disentangling the physical processes contributing to the decline in the overall cosmic star formation rate density ($\rho_\mathrm{SFR}$) since the observed peak at $z\sim$ 2 (e.g. \citealp*{madau2014}) requires the measurement of the gas content in the interstellar medium (ISM) of galaxies out to high redshift. The most reliable technique is to use the neutral hydrogen 21-cm line to trace the atomic gas phase and/or the CO molecule lines arising from rotational transitions to trace the molecular gas component (see e.g. \citealp{carilli2013}, and references therein). However, the linear relationship between the 21-cm line brightness and the column density of gas breaks for optically thick gas (\citealp{braun2009}). Furthermore, the `$\alpha_{\mathrm{CO}}$ factor', the constant of proportionality between the mass of the molecular phase and the CO line emission, typically from the J=1--0 or J=2--1 line, is highly uncertain with a possible dependence on gas-phase metallicity (\citealp{wilson1995}; \citealp{israel2005}), galaxy kinematics, and excitation conditions (\citealp{solomon2005}). The standard CO/21-cm method may also overlook a significant fraction of lower column density molecular gas which is not CO bright and so traced by neither line (\citealp{abdo2010}; \citealp{planck2011}). Technologically, it remains impossible to detect the \hi \ line from galaxies at $z>$0.4 with the current generation of facilities, and the detection of CO line emission typically requires long exposure times (several tens of hours) for normal, high redshift targets.

Faced with these observational difficulties, an alternative to the standard CO/21-cm methods for estimating the mass of the ISM in a galaxy at high redshift might be to use instead the continuum dust emission (see e.g. \citealp{hildebrand1983}; \citealp{dunne2000}; \citealp{boselli2002}). The \textit{Herschel} Space Observatory (\citealp{pilbratt2010}) with the Photodetector Array Camera and Spectrometer (PACS; \citealp{poglitsch2010}) and the Spectral and Photometric Imaging REceiver (SPIRE; \citealp{griffin2010}) were jointly capable of detecting the far-infrared (FIR) to submillimetre (submm) continuum emission originating from the dust component in six wavebands (70 to 500~$\mu$m) with significantly higher sensitivity and angular resolution than previous FIR/submm experiments, making it possible to derive a calibration between the dust emission and the ISM mass, $M_{\rm ISM}$ (\citealp{eales2012}; \citealp{magdis2013}), though the calibration is dependent \mbox{on an accurate knowledge of the dust temperature.}

Most recently, \citet{scoville2016} used a calibration between the dust continuum at $\lambda =$ 850~$\mu$m and the molecular gas content to infer the properties of higher redshift ($z\leq 6$) galaxies. The empirical calibration was obtained considering \textit{Planck} observations of the Milky Way (\citealp{planck2011xxi,planck2011xxv}) and samples of low redshift star-forming galaxies (\citealp{dale2005}; \citealp{clemens2010}), ultraluminous infrared galaxies (ULIRGs), and higher redshift ($z=$2--3) submillimetre galaxies (SMGs) from the literature. Although they report that each method yields a similar rest-frame 850~$\mu$m luminosity per unit ISM mass, the calibration based on the sample of 70 star-forming galaxies, SMGs and ULIRGs, gave $L_{\nu_{850}}$/$M_{\mathrm{H}_{2}}$~$=$~(6.7$\pm$1.7)$\times$10$^{19}$~erg~s$^{-1}$~Hz$^{-1}$~M$_{\odot}^{-1}$. By applying their calibration to ALMA observations of galaxies in three redshift bins up to $z=4.4$, \citeauthor{scoville2016} conclude that  starburst galaxies above the main sequence are largely the result of having greatly increased gas masses rather than an increased efficiency of converting gas to stars, with star-forming galaxies at $z>1$ exhibiting $\sim$2--5 times shorter gas depletion times than low-z galaxies. Whilst the aplication of this empirical calibration (see also \citealp{scoville2017}) has clear advantages, being much faster ($\sim$ 20$\times$) at estimating the ISM mass than molecular line observations and applicable to more readily obtainable continuum observations at higher redshift, the method assumes a solar metallicity and so may not apply to lower mass, metal-poor galaxies at higher redshifts. It is crucial to test the robustness of this calibration to ensure that any evolution \mbox{with redshift is in fact physical.} 

In this Letter, we present the calibration between the dust continuum and molecular gas content derived from measurements of $L_{\nu_{850}}$, $M_{\mathrm{H}_{2}}$ and $M_{\rm ISM}$ for an expanded, homogeneous sample of 67 main-sequence star-forming galaxies at \mbox{$0.02<z<0.35$} in the Valpara\'{i}so ALMA Line Emission Survey (VALES; \citealp{villanuevaprep}; \citealp{hughessub}), based on a combination of Band 3 \co \ observations taken with the Atacama Large Millimetre/submillimetre Array (ALMA) and FUV-submm photometry from the GAMA survey (\citealp{driver2016}; \citealp{wright2016}). We adopt a $\Lambda$CDM cosmology with $H_0=70$\,km\,s$^{-1}$\,Mpc$^{-1}$, \mbox{$\Omega_{\rm M}=0.27$ and $\Omega_{\Lambda}=0.73$}.

\begin{figure}
\begin{center}
\includegraphics[width=0.95\columnwidth]{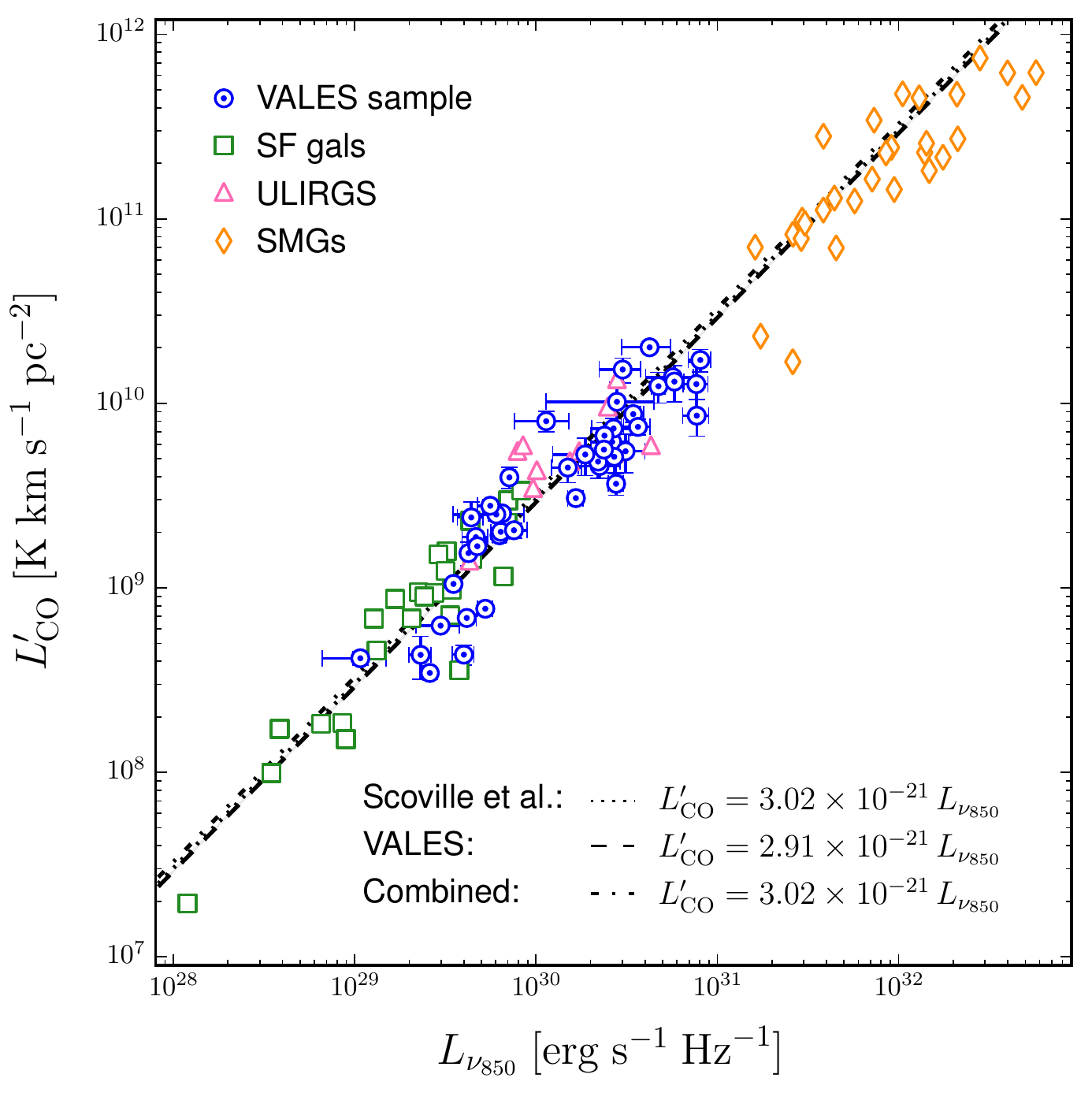}
\end{center}
\vspace{-0.5cm}
\caption{The correlation between $L_{\nu_{850}}$ and ${L}'_{\rm CO}$ found for galaxies in our VALES sample observed with ALMA (\citealp{villanuevaprep}; \citealp{hughessub}). For galaxies with CO detections (blue circles), we show the average ratio (black dashed line) and compare to the mean value (black dotted line) found for the low-$z$ samples of star-forming galaxies (SF; squares), ultraluminous infrared galaxies (ULIRGs; triangles), and submillimetre galaxies (SMGs; diamonds) studied in \citet{scoville2016}. The average of these combined samples is superimposed (dashed-dotted line).}\label{fig:figure1}
\end{figure}

\section{The sample and data}\label{sec:sampledata}

\subsection{Sample selection}

Our sample of galaxies was originally drawn from the \textit{Herschel} Astrophysical Terahertz Large Area Survey (\citealp{eales2010}; \citealp{valiante2016}; \citealp{bourne2016}), a \textit{Herschel} programme capable of providing a sufficient number of far-IR bright galaxies over $\sim$600 deg$^{2}$ with a wealth of high-quality ancillary data. From the three equatorial fields spanning $\sim$160 deg$^{2}$ covered by $H$-ATLAS, galaxies were selected based on the following criteria: (1) a flux of $S_{160 \mu\mathrm{m}} >$ 150 mJy; (2) no neighbours with $S_{160 \mu\mathrm{m}} > 160$~mJy ($3 \sigma$) within 2 arcmin from their centroids; (3) an unambiguous identification (\textsc{reliability} $>$0.8, \citealp{bourne2016}) in the Sloan Digital Sky Survey (SDSS DR7; \citealp{abazajian2009}); (4) a Petrosian SDSS $r$-band radius $<$15$\arcsec$, i.e. smaller than the PACS spectroscopic field of view; (5) high-quality spectroscopic redshifts (\textsc{zqual} $>$ 3) from the Galaxy and Mass Assembly survey (GAMA; \citealp{liske2015}); and (6) a redshift between 0.02 $<z<$ 0.35 (median of 0.05), beyond which the CO(1--0) line is redshifted out of ALMA Band 3. After applying these criteria, 324 galaxies remain to comprise a statistically-significant sample spanning a wide range of optical morphological types and IR luminosities. Of these, 67 objects have follow-up ALMA \co \ line observations as part of VALES, and GAMA FUV to FIR/submm photometry. These galaxies have stellar masses from 6 to 11$\times$10$^{10}$~M$_{\odot}$, SFRs between 0.6 and 100~M$_{\odot}$~yr$^{-1}$, and metallicities of $8.7<12+\log_{10}$(O/H)$<9.2$ (see \citealp{villanuevaprep}).

\subsection{ALMA \co \ line observations}

We exploit our VALES observations targeting the \mbox{\co} \ line in Band 3 for 67 galaxies obtained during cycle-1 and -2. \citet{villanuevaprep} present the observations, data reduction and a detailed characterisation for the complete sample. All observations were reduced homogeneously within the \textit{Common Astronomy Software Applications}\footnote{\url{http://casa.nrao.edu/index.shtml}} (CASA; \citealp{mcmullin2007}) using a common pipeline, developed from standard pipelines, for calibration, concatenation and imaging, with standard bandpass, flux and phase calibrators. Velocity-integrated \co \ flux densities, $S_{\mathrm{CO}} \Delta \rm v$, in units of Jy~km~s$^{-1}$ were obtained by collapsing the cleaned, primary-beam-corrected data cubes between $\nu_{\mathrm{obs}}-\nu_{\mathrm{FWHM}}$ and $\nu_{\mathrm{obs}}+\nu_{\mathrm{FWHM}}$, and fitting these cubes with a Gaussian. We detect $>73$\% (49 of 67) of the targets with a $>5 \sigma$ peak line detection. We estimate upper limits as 5$\times$ the measured RMS from the collapsed cubes set at 100~km~s$^{-1}$ spectral resolution and adopting $\nu_{\mathrm{FWHM}}$ = 250~km~s$^{-1}$.

\subsection{GAMA multiwavelength photometry}

All of our galaxies are present in the GAMA Panchromatic Data Release\footnote{\url{http://cutout.icrar.org/panchromaticDR.php}} (\citealp{driver2016}) that provides imaging for over 230 deg$^{2}$ with photometry in 21 bands extending from the far-ultraviolet to far-infrared from numerous facilities, currently including: GALaxy Evolution eXplorer (GALEX), Sloan Digital Sky Survey (SDSS), Visible and Infrared Telescope for Astronomy (VISTA), Wide-field Infrared Survey Explorer (WISE), and \textit{Herschel}. These data are processed to a common astrometric solution from which homogeneous photometry is derived for $\sim$221 373 galaxies with $r<$19.8 mag (see \citealp{wright2016}), meaning the spectral energy distribution (SED) between 0.1--500~$\mu$m is \mbox{available for each galaxy.}

\section{The $L_{\nu_{850}}$--$M_{\mathrm{ISM}}$ calibration}\label{sec:calibration}

\subsection{Estimating the dust continuum luminosity}\label{sec:magphys}

In the absence of measurements of the dust continuum at 850 $\mu$m, we adopt an estimate of the $L_{\nu_{850}}$ based on an extrapolation of the modelled SED. Our primary approach to estimate $L_{\nu_{850}}$ exploits the FUV--FIR/submm \textit{H}-ATLAS/GAMA photometry available for all our galaxies modelled with the Bayesian SED fitting code, MAGPHYS \citep{dacunha2008}. The code fits the panchromatic SED, giving special consideration to the dust--energy balance, from a library of optical and infrared SEDs derived from a generalised multi-component model of a galaxy. The FIR/submm dust emission is modelled with five modified black bodies, of which two components have variable temperatures representing thermalised cold and warm dust and the other three components represent hot dust at 130, 250 and 850~K. Two geometries describe the dust distribution: birth clouds of new stars contain only warm and hot circumstellar dust, whereas all five dust components may contribute to the dust in the diffuse ISM. As our focus is solely on estimating the rest frame 850~$\mu$m continuum luminosity, we refer the reader to  \citet{driverprep} for details of the complete analysis of the MAGPHYS modelling of all the GAMA SEDs, yet note that \citet{villanuevaprep} demonstrate how the stellar masses, IR luminosities, $L_{\mathrm{IR}}$, and SFRs derived from MAGPHYS are consistent within the uncertainties to empirical estimates found in \citet{ibar2015}.

Using the best-fit SEDs, we calculate the median model flux between 800 and 900~$\mu$m, $S_{\nu_{850}}$, and convert this flux -- that ranges from 1 to 15 mJy -- into a monochromatic rest-frame luminosity, $L_{\nu_{850}}$, in units of $\mathrm{erg}~\mathrm{s}^{-1}~\mathrm{Hz}^{-1}$, via
\begin{equation}
L_{\nu_{850}} = 1.19\times10^{27}\,S_{\nu_{850}}(\mathrm{Jy})\,(1+z)^{-1}\,D^{2}_{L}\,K~\mathrm{erg}~\mathrm{s}^{-1}~\mathrm{Hz}^{-1}
\end{equation}
where $D_{\mathrm{L}}$ is the luminosity distance in Mpc, and $K$ is the $K$-correction given by Eqn.~2 in \citet{dunne2011}, following their exact methodology\footnote{The mean scatter between $L_{\nu_{850}}$ calculated via the method presented in Appendix A of \citet{scoville2016} and that used here is $\pm$5\%, and both methods yield similar scaling relations.}. In addition to FUV--FIR/submm SED modelling via MAGPHYS, we also examine the results of fitting the five \textit{H}-ATLAS PACS/SPIRE photometric bands with a one-component modified blackbody as originally presented by \citet{hildebrand1983}, assuming a power-law dust emissivity and either keeping the spectral index $\beta$ as a free parameter or fixing the value at 1.8 (e.g. \citealp{galametz2012}). In both cases, our best-fit model fluxes are consistent and produce results that support the conclusions reached with the MAGPHYS SED fitting results. We then compute the uncertainty in $L_{\nu_{850}}$ from the standard deviation of the three luminosity values we obtain from modelling the SEDs with MAGPHYS and the two fits with one-component modified black bodies adopting variable and fixed spectral indices.  
	
\begin{figure}
\begin{center}
\includegraphics[width=0.95\columnwidth]{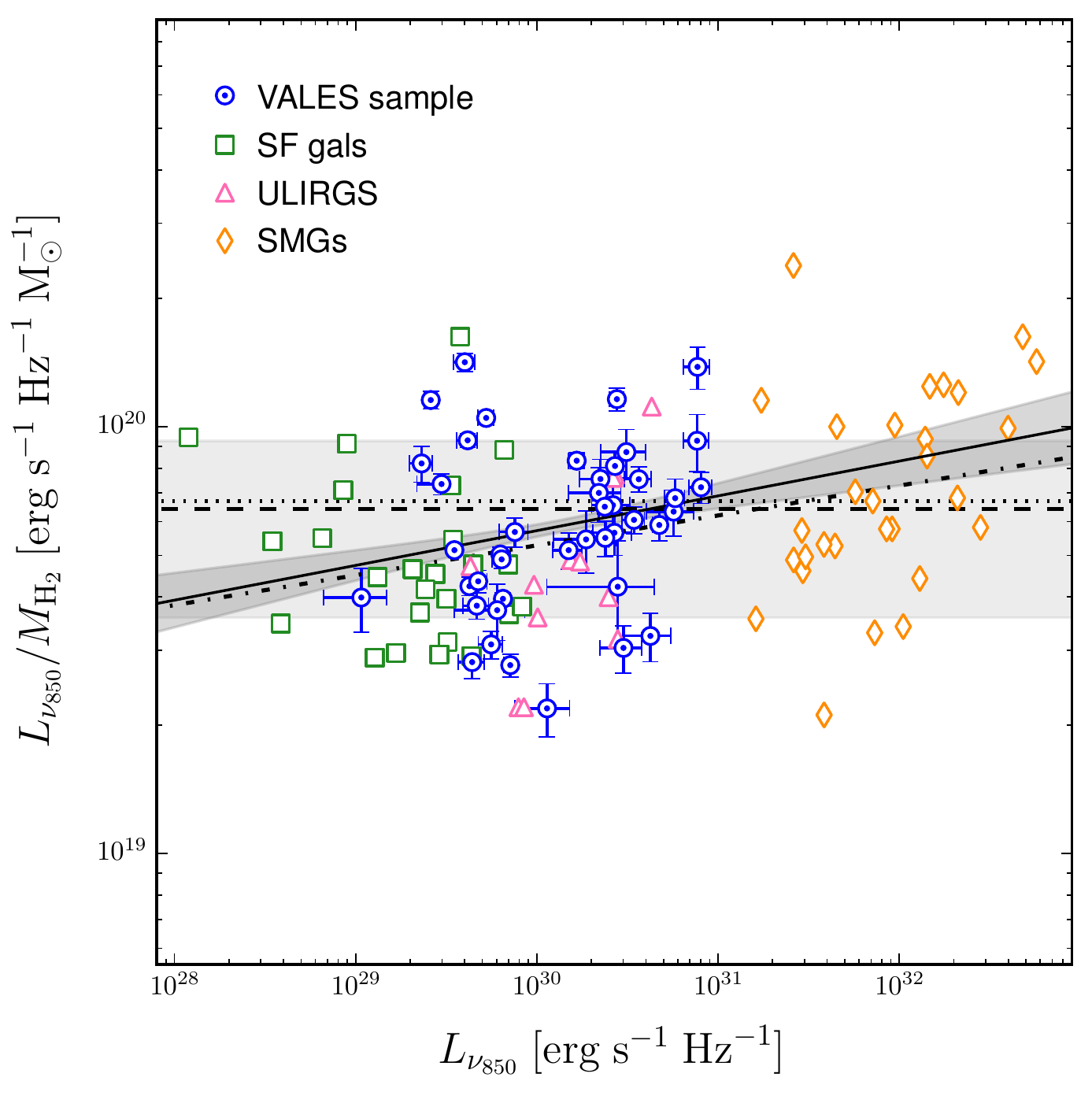}
\end{center}
\vspace{-0.5cm}
\caption{The ratio of $L_{\nu_{850}}$ to $M_{\mathrm{H}_{2}}$ found for galaxies in our VALES sample observed with ALMA (\citealp{villanuevaprep}; \citealp{hughessub}). From galaxies with CO detections (blue circles), we show the average ratio (black dashed line) with the $\pm1\sigma$ range (light grey region) and the best linear fit (black solid line) with $\pm1\sigma$ confidence limits (dark grey region). The low-$z$ samples of SF galaxies (squares), ULIRGs (triangles), and SMGs (diamonds) are shown together with the mean value (black dotted line), taken from Fig.~1 of \citet{scoville2016}.}\label{fig:figure2}
\end{figure}

\subsection{Measurement of the interstellar gas content}

From our velocity-integrated \co \ flux densities, $S_{\mathrm{CO}} \Delta \rm v$, in units of Jy~km~s$^{-1}$, we calculate the CO line luminosity, ${L}'_{\rm CO}$, in units of ${\rm K\,km\,s^{-1}\,pc^{2}}$ following Eqn.~3 of \citealp*{solomon2005}, given as
\begin{equation}
{L}'_{\rm CO} = 3.25 \times 10^{7}\, S_{\rm CO} \, \Delta \rm v \, \nu^{-2}_{\rm obs} \, D^{2}_{\mathrm{L}} \, (1+z)^{-3}, 
\end{equation}
where $\nu_{\rm obs}$ is the observed frequency of the emission line in GHz. The values for ${L}'_{\rm CO}$ are in the range of ($0.03 - 3.51$) $\times \,10^{10}$\, ${\rm K\,km\,s^{-1}}$\,pc$^{2}$, with an average value of ($0.67 \pm 0.06$)~$\times ~10^{10}$\, ${\rm K\,km\,s^{-1}}$\,pc$^{2}$. The CO line luminosity can then be converted into the molecular gas mass (including the mass of He), $M_{\rm H_{2}}$, by assuming an $\alpha_{\rm CO}$ conversion factor \citep[see Eqn.~5 in ][]{solomon2005}. Our VALES galaxies have high stellar masses ($\geq10^{10}$~M$_{\odot}$), thus avoiding metal-poor systems in which the dust-to-gas abundance ratio is expected to decrease nor where significant molecular gas exists without CO emission (see e.g. \citealp{bolatto2013}). 

We first exclude from our analysis the merger/interacting systems identified using a K-band-based morphological classification as outlined in \citet{villanuevaprep}. To facilitate a direct comparison with the results of \citeauthor{scoville2016}, we primarily adopt $\alpha_{\rm CO}$\,=\,6.5\,(K\,km\,s$^{-1}$\,pc$^{2}$)$^{-1}$ for the bulge- and disk-dominated galaxies with normal star formation. In our 43 normal galaxies with detected CO emission, we derive $M_{\rm H_{2}}$ values in the range of $\log{M_{\rm H_{2}}/ \rm M_{\odot}} = 9.35-11.12$ with an median of $10.46 \pm 0.01$. Finally, to estimate the atomic hydrogen content, $M_{\mathrm{H}{\textsc{i}}}$, we use the \hi --colour scaling relation given by Eqn. 4 of \citet{zhang2009} with the $g-r$ colour and  $i$--band surface brightness available from the GAMA photometry. The \hi \ mass ranges between $\log M_{\mathrm{H}{\textsc{i}}}/ \mathrm{M}_{\odot} = 8.55-10.54$ with an average of $9.57 \pm 0.03$ and typical errors of $\pm$30\%. We then calculate the ISM gas mass as $M_{\rm ISM} = M_{\mathrm{H {\textsc{i}}}} + M_{\rm H_{2}}$ using standard error propagation.

\subsection{The $L_{\nu_{850}}$/$M_{\mathrm{ISM}}$ calibration}

Bringing these measurements together, we now examine the calibrations between the dust continuum and gas content found for the galaxies in our VALES sample considering only those galaxies with CO line detections. The VALES sample exhibits a mean ratio $S_{\mathrm{CO}} \Delta \rm v$/$S_{\nu_{850}}$ of 1081$\pm$265~km~s$^{-1}$, corrresponding to a mean ${L}'_{\rm CO}$ to $L_{\nu_{850}}$ ratio of (2.91$\pm$0.66)$\times$10$^{-21}$ in units of the luminosity ratio dimensions (see \hyperref[fig:figure1]{Fig.~\ref*{fig:figure1}}), which is in agreement with that found (3.02$\times$10$^{-21}$) for the three galaxy samples analysed in \citet{scoville2016}. In particular, the VALES galaxies have properties more akin to the low-$z$ normal star-forming galaxies and ULIRGs than the SMG sample. After converting the CO luminosity into molecular gas mass, we find average ratios of $S_{\nu_{850}}$/$M_{\mathrm{H}_{2}}$~$=$~(6.9$\pm$5.6)$\times$10$^{-13}$~Jy~M$_{\odot}^{-1}$ and $L_{\nu_{850}}$/$M_{\mathrm{H}_{2}}$~$=$~(6.4$\pm$1.4)$\times$10$^{19}$~erg~s$^{-1}$~Hz$^{-1}$~M$_{\odot}^{-1}$, also in excellent agreement with the mean values found by \citet{scoville2016} and with near-matching scatter.

Although a constant ratio is appropriate to describe the average properties of the both the \citeauthor{scoville2016} and VALES samples across the luminosity range, there is a very minor trend that galaxies with $L_{\nu_{850}}>10^{30}$~erg~s$^{-1}$~Hz$^{-1}$ tend to lie on or above the average ratio (see \hyperref[fig:figure2]{Fig.~\ref*{fig:figure2}}). Galaxies with $L_{\nu_{850}}$ fainter than this luminosity have slightly lower ratios than the average. Our results suggest that adopting a constant $L_{\nu_{850}}$/$M_{\mathrm{H}_{2}}$ ratio to estimate the ISM mass would underestimate $M_{\mathrm{H}_{2}}$ in galaxies where $L_{\nu_{850}}>10^{30}$~erg~s$^{-1}$~Hz$^{-1}$ (and vice versa) and so a linear fit (in logarithmic space) may be more appropriate for the  galaxies. For the VALES sample, we obtain
\begin{align}
 & \log_{10} {M}_{\mathrm{H}_{2}}& =\,  (0.92\pm0.02) \log_{10} L_{\nu_{850}}-(17.31\pm0.59), \\
 & \log_{10} {M}_{\mathrm{ISM}}& =\,  (0.86\pm0.02) \log_{10} L_{\nu_{850}}-(15.38\pm0.55), \label{eqn:fit}
\end{align}
in which ${M}$ and $L_{\nu_{850}}$ have units of ${\mathrm{M}_{\odot}}$ and ${\mathrm{erg}~\mathrm{s}^{-1}~\mathrm{Hz}^{-1}}$, respectively. This relation is valid between $1\times 10^{29}<L_{\nu_{850}}<2\times 10^{31}$~${\mathrm{erg}~\mathrm{s}^{-1}~\mathrm{Hz}^{-1}}$ for normal main-sequence star-forming galaxies and is based on the assumption that $\alpha_{\rm CO}$\,=\,6.5\,(K\,km\,s$^{-1}$\,pc$^{2}$)$^{-1}$. Furthermore, we consider the calibrations we obtain from combining the 70 galaxies of \citet{scoville2016} with the 43 CO-detected star-forming galaxies in our VALES sample. The $L_{\nu_{850}}$--$M_{\mathrm{H}_{2}}$ calibration for this combined sample of 113 objects is then
\begin{align}
 & \log_{10} {M}_{\mathrm{H}_{2}}& =\,  (0.93\pm0.01) \log_{10} L_{\nu_{850}}-(17.74\pm0.05)
\end{align}
with a scatter of $\sim$0.1 dex. However, the dominant error is on $\alpha_{\rm CO}$ and is not included in our error calculations. We note that although \citet{scoville2014} include the \hi \ mass contribution to $M_\mathrm{ISM}$ (estimated as 50\% of the H$_{2}$ mass), \citet{scoville2016} only consider the H$_{2}$ mass component, therefore we do not include a $L_{\nu_{850}}$--$M_{\mathrm{ISM}}$ calibration for the combined sample. We summarise these best-fit relations and the corresponding correlation coefficients in \hyperref[tab:table1]{Table~\ref*{tab:table1}}, in which we also present the relations we obtain when adopting the Galactic value of $\alpha_{\rm CO}$\,=\,4.6\,(K\,km\,s$^{-1}$\,pc$^{2}$)$^{-1}$ for our sample.

\section{Discussion}\label{sec:discussion}

We have reported an updated calibration between the dust continuum and molecular gas content for an expanded sample of 67 main-sequence star-forming galaxies at $0.02<z<0.35$ drawn from the \textit{H}-ATLAS, using gas mass measurements from ALMA Band-3 \co \ observations and estimates of the monochromatic luminosity at 850~$\mu$m (rest-frame), $L_{\nu_{850}}$, via an extrapolation of the dust continuum from MAGPHYS modelling of the FUV to FIR/submm SED observed by the GAMA survey. Although we confirm an average $L_{\nu_{850}}$/$M_{\mathrm{H}_{2}}$ ratio in close agreement with literature values (see \citealp{scoville2016}, and references therein), the linear fit given by \hyperref[eqn:fit]{Eqn.~\ref*{eqn:fit}} alleviates the issue that the ISM mass may be overestimated for galaxies with lower continuum luminosities. Whilst we recommend using this best-fit calibration rather than the constant calibration for estimating the gas content from dust continuum observations of main-sequence galaxies at high redshift, we stress that the largest uncertainty in this work remains in the $\alpha_{\rm CO}$ factor.

Whichever value of $\alpha_{\rm CO}$ we choose to adopt, using these Galactic-type values assumes that the CO emission comes from viriliazied molecular clouds bound by self-gravity. However, it remains possible that the CO line emission may actually trace material bound by the total potential of the galactic center consisting of a mass of stars and dense gas clumps equal to the dynamical mass, $M_\mathrm{dyn}$, and a diffuse interclump medium the CO emitting gas of mass $M_\mathrm{gas}$. In this case, $M_\mathrm{gas} = M_\mathrm{dyn} (\alpha_{\rm CO}\,{L}'_{\rm CO})^{2}$ (see Eqn. 6 in \citealp*{solomon2005}), meaning the usual relation of $\alpha_{\rm CO}$ will be changed if a fraction of the CO emission in our galaxies originates from an intercloud medium bound by the galaxy potential. Unfortunately, we currently have too few normal disk-dominated galaxies with spatially-resolved CO emission (7 sources in total) to robustly identify whether such a change to the $\alpha_{\rm CO}$ factor is warranted in our sample. We would also require more accurate estimates covering a greater dynamic range of metallicity in order to test the effect of the $\alpha_{\rm CO}$ dependency on metallicity. In future VALES studies, we aim to use ALMA and MUSE to further investigate the robustness of the calibration between the dust continuum and molecular gas content with an $\alpha_{\rm CO}$ constrained by 3D kinematical modelling for a larger sample of resolved galaxies.

\renewcommand{\arraystretch}{1.0}
\begin{table}
 \centering
 \begin{minipage}{\columnwidth}
  \caption{The best-fit relations between the dust continuum luminosity at 850~$\mu$m (in units of erg~s$^{-1}$~Hz$^{-1}$) and various parameters considered in this work, expressed as $\log_{10} y =m \log_{10}L_{\nu_{850}}+c$, with the corresponding 1$\sigma$ errors. We also state the scatter in dex ($\sigma$), Spearman ($\rho_\mathrm{S}$), and Pearson ($\rho_\mathrm{P}$) coefficients, where the values have probabilities P($\rho_\mathrm{P}$) and P($\rho_\mathrm{S}$) of $>$99.9\% that each of the two variables correlate for \mbox{sample size $N=43$ or 113.} $^{\color{blue} a}$Combined sample of 113 galaxies in VALES and \citet{scoville2016}. \vspace{-0.2in}}\label{tab:table1}
\begin{center}
\begin{tabular}{l c c c c c}
\hline
\hline
$y$ & $m$ & $c$ & $\sigma$ & $\rho_\mathrm{P}$ & $\rho_\mathrm{S}$\\
\hline
\multicolumn{6}{l}{VALES sample only; $\alpha_{\rm CO}\,\equiv$\,4.6\,(K\,km\,s$^{-1}$\,pc$^{2}$)$^{-1}$}\\
\hline
${L}^{\prime}_{\rm CO}$           & 0.92 $\pm$ 0.02 & -18.13 $\pm$ 0.59 & 0.12 & 0.94 & 0.92 \\
$M_{\mathrm{H}_{2}}$/$M_{\odot}$  & 0.92 $\pm$ 0.02 & -17.46 $\pm$ 0.59 & 0.12 & 0.94 & 0.92 \\
$M_{\mathrm{ISM}}$/$M_{\odot}$    & 0.84 $\pm$ 0.02 & -14.95 $\pm$ 0.54 & 0.11 & 0.94 & 0.92 \\
\hline
\multicolumn{6}{l}{Combined samples$^{\color{blue} a}$; $\alpha_{\rm CO}\,\equiv$\,4.6\,(K\,km\,s$^{-1}$\,pc$^{2}$)$^{-1}$}\\
\hline
${L}^{\prime}_{\rm CO}$           & 0.93 $\pm$ 0.01 & -18.56 $\pm$ 0.05 & 0.12 & 0.98 & 0.98 \\
$M_{\mathrm{H}_{2}}$/$M_{\odot}$  & 0.93 $\pm$ 0.01 & -17.90 $\pm$ 0.05 & 0.12 & 0.98 & 0.98 \\
\hline
\multicolumn{6}{l}{VALES sample only; $\alpha_{\rm CO}\,\equiv$\,6.5\,(K\,km\,s$^{-1}$\,pc$^{2}$)$^{-1}$}\\
\hline
${L}^{\prime}_{\rm CO}$           & 0.92 $\pm$ 0.02 & -18.13 $\pm$ 0.59 & 0.12 & 0.94 & 0.92 \\
$M_{\mathrm{H}_{2}}$/$M_{\odot}$  & 0.92 $\pm$ 0.02 & -17.31 $\pm$ 0.59 & 0.12 & 0.94 & 0.92 \\
$M_{\mathrm{ISM}}$/$M_{\odot}$    & 0.86 $\pm$ 0.02 & -15.38 $\pm$ 0.55 & 0.12 & 0.94 & 0.92 \\
\hline
\multicolumn{6}{l}{Combined samples$^{\color{blue} a}$; $\alpha_{\rm CO}\,\equiv$\,6.5\,(K\,km\,s$^{-1}$\,pc$^{2}$)$^{-1}$}\\
\hline
${L}^{\prime}_{\rm CO}$           & 0.93 $\pm$ 0.01 & -18.56 $\pm$ 0.05 & 0.12 & 0.98 & 0.98 \\
$M_{\mathrm{H}_{2}}$/$M_{\odot}$  & 0.93 $\pm$ 0.01 & -17.74 $\pm$ 0.05 & 0.12 & 0.98 & 0.98 \\
\hline
\end{tabular}
\end{center}
\end{minipage}
\end{table}

\section*{Acknowledgments} 
\small
TMH and EI acknowledge CONICYT/ALMA funding Program in Astronomy/PCI Project N$^\circ$:31140020. MA acknowledges partial support from FONDECYT through grant 1140099. DR acknowledges support from the National Science Foundation under grant number AST-1614213 to Cornell University. LD, SJM and RJI acknowledge support from European Research Council Advanced Investigator Grant COSMICISM, 321302; SJM and LD are also supported by the European Research Council Consolidator Grant {\sc CosmicDust} (ERC-2014-CoG-647939). YQX acknowledges support from grants NSFC-11473026 and NSFC-11421303. This paper uses the following ALMA data: ADS/JAO.ALMA \#2012.1.01080.S \& \#2013.1.00530.S. ALMA is a partnership of ESO (representing its member states), NSF (USA) and NINS (Japan), together with NRC (Canada), NSC and ASIAA (Taiwan), and KASI (Republic of Korea), in cooperation with the Republic of Chile. The Joint ALMA Observatory is operated by ESO, AUI/NRAO and NAOJ.



\bibliographystyle{mnras}
\bibliography{mn170523l} 

\bsp	
\label{lastpage}
\end{document}